\documentclass[12pt]{article}
\usepackage{amsmath,amssymb,amsthm}
\usepackage{latexsym}

\newtheorem{lemma}{Lemma}
\newtheorem{theorem}{Theorem}

\begin{document}

\title{Quadratic algebras and integrable chains}
\author{Fedor Soloviev\\
Courant Institute of Mathematical Sciences,\\
251 Mercer St, New York, NY 10012\\
soloviev@nyu.edu
}
\maketitle

\begin{abstract}
Using Krichever-Phong's universal formula, we show that a multiplicative representation linearizes Sklyanin
quadratic brackets for a multi-pole Lax function with a spectral parameter. The spectral parameter can be either
rational or elliptic. As a by-product, we obtain an extension of a Sklyanin algebra in the elliptic case.
Krichever-Phong's formula provides a hierarchy of symplectic structures, and we show that there exists
a non-trivial cubic bracket in Sklyanin's case.
\end{abstract}

\section{Introduction}

A starting point for many soliton systems is a Lax equation
\begin{equation}\label{Lax-eq}
\dot{L} = [P, L],
\end{equation}
where $L$ and $P$ are operators.

For most finite-dimensional integrable systems, this equation can
be interpreted as a flow on the space $\mathcal{L}$ of meromorphic matrix
functions $L(z)$ on some Riemann surface $\Gamma$ (typically, a Riemann sphere or an elliptic curve), where
the positions of the poles are fixed.

The algebraic-geometric procedure for constructing the exact
solutions (see~\cite{IK02} for a brief outline and additional references) is merely a parametrization
of the space $\mathcal{L}$ in terms of a spectral curve with marked points and a divisor.

A spectral curve is defined by the equation
\begin{equation}\label{speccurve}
\hat{\Gamma}: R(k, z) = \text{det} \thinspace (L(z)-k I)=0.
\end{equation}

For $L(z)$ in general position, it is a smooth Riemann surface of finite
genus. For every point $Q=(k,z)$ of $\hat{\Gamma}$, there exists
the unique eigenvector $\psi(Q)$ of $L(z)$ satisfying
$$
L(z) \psi(Q) = k \psi(Q),
$$
normalized so that its first component is one, i.e.,
$\psi_1(Q) \equiv 1$. The vector function $\psi(Q)$ is meromorphic
on $\hat{\Gamma}$. Due to the Lax equation, the spectral curve
does not change with time and the equivalence class of the pole
divisor of $\psi(Q)$ evolves linearly on the Jacobian of
$\hat{\Gamma}$. The algebraic-geometric procedure allows us to
construct $\psi(Q)$ and $L(z)$ explicitly in terms of Riemann
theta-functions, given an appropriate Riemann surface and a
divisor on it. This procedure does not require any Hamiltonian
theory, although, in most cases, the corresponding physical systems
are governed by completely integrable Hamiltonian equations.

Krichever and Phong~\cite{KP97,KP98} suggested a general approach
to the Hamiltonian theory of integrable systems with Lax-type
equations. They introduced a two-form on the space
$\mathcal{L}$ representing a Hamiltonian structure of the
system. Their formula is universal and works even in the
infinite-dimensional case. In our case, it is defined as:
\begin{equation}\label{KP-uform}
\omega_n = -\dfrac{1}{2} \sum_q \underset{q}{\text{res}} \thinspace \text{Tr}\left( \Psi^{-1}
L^{1-n} \delta L \wedge \delta \Psi - \Psi^{-1} \delta \Psi \wedge K^{1-n} \delta K \right) dz =
-\dfrac{1}{2}\sum_q \underset{q}{\text{res}} \thinspace \Omega dz,
\end{equation}
where $\Psi$ is an eigen-matrix of $L(z)$, i.e., $L \Psi = \Psi K$
and its columns are just vectors $\psi(Q)$ on different sheets
of $\hat{\Gamma}$. The sum is taken over the poles of $L(z)$, the zeroes of $\text{det} \thinspace L(z)$, and
the poles of $dz$ (if there are any). The number $n$ is an integer parameter.
Formula~(\ref{KP-uform}) is well-defined on the space $\mathcal{L}$, but
it depends on the normalization of $\psi(Q)$. It becomes independent of the normalization
when restricted to the leaves, where the one-form $k^{1-n} \delta k dz$ is holomorphic. As a by-product,
$\omega_n$ becomes non-degenerate and independent on gauge
transformations $L \to g L g^{-1}$ on the leaves (for the proofs see~\cite{IK02} and~\cite{IK04}).

An alternative approach to the Hamiltonian theory of integrable
systems uses a so-called r-matrix. An r-matrix defines Poisson
brackets on $\mathcal{L}$ between $L(u)$ and $L(v)$ for fixed $u$
and $v$. In particular, when the domain of $L(z)$ is a Riemann sphere, then, the simplest non-trivial r-matrix is
$$
r(z)=\dfrac{1}{z} \sum_{i,j=1}^N e_{ij} \otimes e_{ji} = \dfrac{\mathcal{P}}{z}.
$$
We denote the domain of $L(z)$ by $\Gamma$ , and call $z$ a spectral parameter.
When $L(z)$ is a $2 \times 2$ (so $N=2$) matrix function, an elliptic r-matrix (i.e., $\Gamma$ is an elliptic curve) is
equal (up to a scalar factor) to~(\cite{Sk79}):
\begin{equation}\label{r-elliptic}
r(z) = -\dfrac{1}{2 \pi} \dfrac{\theta'_{11}}{\theta_{11}(z)} \left( \dfrac{\theta_{01}(z)}{\theta_{01}} \sigma_1 \otimes \sigma_1 +
\dfrac{\theta_{00}(z)}{\theta_{00}} \sigma_2 \otimes \sigma_2 +
\dfrac{\theta_{10}(z)}{\theta_{10}} \sigma_3 \otimes \sigma_3 \right),
\end{equation}
where $\sigma_i$ are the Pauli matrices
$$
\sigma_1=\begin{pmatrix} 0 & 1\\1 & 0 \end{pmatrix},
\sigma_2=\begin{pmatrix} 0 & -\imath\\ \imath & 0 \end{pmatrix},
\sigma_3=\begin{pmatrix} 1 & 0\\0 & -1 \end{pmatrix}.
$$
The notation for the Jacobi theta-functions $\theta_{ij}(z)$ is the same as in Mumford~\cite{M83}.
Notice, that in the $2 \times 2$ case (here $\sigma_0$ is the identity matrix):
$$
\mathcal{P} = \dfrac{1}{2} \left( \sigma_0 \otimes \sigma_0 + \sigma_1 \otimes \sigma_1 + \sigma_2 \otimes \sigma_2 + \sigma_3 \otimes \sigma_3 \right).
$$
These r-matrices satisfy the classical Yang-Baxter equation
$$
[r_{12}(u-v), r_{13}(u)] + [r_{12}(u-v), r_{23}(v)] + [r_{13}(u), r_{23}(v)]=0,
$$
which holds in the space $C^{N^2} \otimes C^{N^2} \otimes C^{N^2}$ with the evident notation $r_{12} = r \otimes I$, $r_{23} = I \otimes r$, etc.

For each r-matrix, one can construct two types of Poisson brackets: a linear bracket and a quadratic bracket.
A linear (or Lie algebraic) bracket is defined as~(\cite{Sk79})
\begin{equation}\label{R-linear}
\{ L(u) \stackrel{\otimes}{,} L(v) \}_1 = [r(u-v), L(u) \otimes I + I \otimes L(v)],
\end{equation}
and a quadratic (or Lie group) bracket is
\begin{equation}\label{R-quadratic}
\{ L(u) \stackrel{\otimes}{,} L(v) \}_2 = [r(u-v), L(u) \otimes L(v)].
\end{equation}
These Poisson structures and the two-forms $\omega_n$ are degenerate on the whole phase space $\mathcal{L}$.
However, they naturally foliate the space into symplectic manifolds, where it is possible to compare them.

The classical Yang-Baxter equation is a sufficient condition to ensure
that the brackets above are indeed Poisson.
There is no universal way to construct solutions to this equation, although many solutions are known~(\cite{RS94}).

$\omega_1$, given by~(\ref{KP-uform}), coincides with the linear brackets. It is
possible to show~(\cite{IK02}) that the two-form $\omega_1$
coincides with a Kirillov-Kostant form on the direct product of
coadjoint orbits of the $GL(N)$ action. When the poles of $L(z)$ are
simple, the corresponding principal parts can be identified with
Lie algebra duals $\mathfrak{gl}^*(N)$. Symplectic leaves are
determined by the condition that the one-form $\delta K dz$ is
holomorphic, or, equivalently, that the principal parts of $K$ at each
pole of $K$ are fixed. The latter condition fixes some orbits in $\mathfrak{gl}^*(N)$.

It is shown in~\cite{IK04} that quadratic brackets coincide
with $\omega_2$. The nature of quadratic brackets is more
complicated. Quadratic brackets in the elliptic case
were explicitly computed only recently in~\cite{Z07} for a multi-pole Lax function.
Formulas (3.2)-(3.6) in~\cite{Z07} appear complicated with no apparent pattern.
In this paper we show that if one uses a multiplicative representation for
a multi-pole Lax function, then the quadratic bracket assumes a remarkably simple form.
In the case considered in~\cite{Z07}, the bracket may be obtained as a reduction from the direct
product of spaces with single-pole quadratic brackets.

The simplest illustration of these ideas is a rational case, i.e.,
when $\Gamma$ is a Riemann sphere and $L(z)$ is just a meromorphic matrix function with a fixed
number of poles on the extended complex plane.
We assume that $L(z)$ is in ``general position,'' which means that it belongs to a big open
cell in the space of meromorphic functions. In particular, $L(z)$ may only have simple
poles with residues of rank one, and it is diagonalizable at least at one point.
Without loss of generality, we assume that it is diagonalizable at $z=\infty$.
Other cases, when $L(z)$ has a higher-order pole or a residue
of a higher rank, are special and may be obtained as a result of some limiting procedure.

The most natural way to write a meromorphic function with simple poles is to specify positions of
its poles and their corresponding residues. This leads to the formula
$$
L(z) = L_0 + \sum_{i=1}^d \dfrac{a_i b_i^T}{z-z_i},
$$
where $a_i$, $b_i$ are $N$-dimensional vectors and
$L_0$ is a constant matrix. We call it an ``additive representation,'' because it reflects
the additive Lie algebraic structure on the space of Lax functions $\mathcal{L}$.
This representation is well-suited to the linear Poisson bracket (or the corresponding
symplectic form $\omega_1$), which, in these coordinates, equals
$$
\omega_1 = \sum_{i=1}^d \delta a_i^T \wedge \delta b_i.
$$
However, different coordinates are natural for the quadratic bracket.
Any function $L(z)$ has an \emph{equivalent} ``multiplicative representation:''
$$
L= L_0 \left( I + \dfrac{p_1 q_1^T}{z-z_1} \right) \left( I +
\dfrac{p_2 q_2^T}{z-z_2} \right).. \left( I + \dfrac{p_d
q_d^T}{z-z_d} \right),
$$
where $p_i$ and $q_i$ are also $N$-dimensional vectors.

It seems that the coordinates $p_i,q_i$ appear first in~\cite{KK97}, and
later they are used by Borodin~\cite{B04} in the theory of difference equations.
These coordinates emphasize the multiplicative (or ``group'') structure on the space
of Lax operators.
Essentially, a multiplicative representation is a particular case of an ``integrable chain'' from~\cite{IK04}.
However, in~\cite{IK04} integrable chains were not related to an additive representation, and it seems that
the name ``multiplicative representation'' is more appropriate than ``chain'' in the present situation.
The quadratic 2-form corresponding to the quadratic Poisson brackets equals
\begin{equation}\label{rat-Sk}
\omega_2 = \sum_{i=1}^d \delta p_i^T \wedge \delta q_i.
\end{equation}
Notice that it is a highly non-trivial task to arrive at explicit commutation relations
between coordinates $p_i, q_i$ (or, $a_i$ and $b_i$ in the linear case) starting from
Formulas~(\ref{R-linear}) or~(\ref{R-quadratic}). One advantage of Krichever-Phong's
Formula~(\ref{KP-uform}) is that it allows us to find them and that it works equally well
in rational and elliptic cases. A multiplicative representation also exists in the elliptic
case, and similar statements related to the quadratic bracket hold (see Sections 4-6).
We consider separately the case of ``general position,'' when the poles of $L(z)$ are simple and
have rank one and the Sklyanin case of higher rank poles.

It turns out, that for Sklyanin's Lax Matrix~\cite{Sk82}, the 2-form $\omega_2$ given by Formula~(\ref{KP-uform})
is degenerate even on the leaves where $\delta \ln{k} dz$ is holomorphic.
In order to circumvent this difficulty, we need to introduce an additional parameter $u$ to 4 Sklyanin's
variables $s_0,s_1,s_2,s_3$. Surprisingly, this provides an extension of a quadratic Poisson algebra:
$$
\{s_0,s_1\}_2=-\theta_{01}^4 s_2 s_3, \quad
\{s_0,s_2\}_2=\theta_{00}^4 s_1 s_3, \quad
\{s_0,s_3\}_2=-\theta_{10}^4 s_1 s_2,
$$
$$
\{s_1, s_2\}_2 = -s_0 s_3, \quad
\{s_1, s_3\}_2 = s_0 s_2, \quad
\{s_2, s_3\}_2 = -s_0 s_1,
$$
$$
\{u,s_0\}_2 = s_0, \quad
\{u,s_1\}_2 = s_1, \quad
\{u,s_2\}_2 = s_2, \quad
\{u,s_3\}_2 = s_3.
$$
These brackets coincide with the Sklyanin Brackets~\cite{Sk82} without the last 4 identities.
However, the symplectic leaves for the original Sklyanin brackets have dimension 2, whereas
for our extension the leaves have dimension 4. The leaves are determined by one condition
$$
\delta ((s_0^2 + s_1^2 \theta_{00}^4 + s_2^2 \theta_{01}^4)/(s_1^2+s_2^2+s_3^2))=0.
$$

As we said earlier, Formula~(\ref{KP-uform}) provides an hierarchy of 2-forms on the space of
Lax operators $\mathcal{L}$. Different integer values of $n$ correspond to different 2-forms.
In particular, it provides a non-trivial cubic bracket on the space of Sklyanin's Lax functions.
The cubic bracket corresponds to $\omega_3$ and is given by relations
$$
\{s_0,s_1\}_3=-2\theta_{01}^4 s_0 s_2 s_3, \quad
\{s_0,s_2\}_3=2 \theta_{00}^4 s_0 s_1 s_3, \quad
\{s_0,s_3\}_3=-2\theta_{10}^4 s_0 s_1 s_2,
$$
$$
\{s_1, s_2\}_3 = s_3(s_1^2 \theta_{00}^4 + s_2^2 \theta_{01}^4 - s_0^2), \quad
\{s_1, s_3\}_3 = s_2(s_0^2 - s_1^2 \theta_{10}^4 + s_3^2 \theta_{01}^4),
$$
$$
\{s_2, s_3\}_3 = -s_1(s_0^2 + s_2^2 \theta_{10}^4 + s_3^2 \theta_{00}^4).
$$
This bracket is non-degenerate on the leaves
$\delta u = \delta ((s_0^2 + s_1^2 \theta_{00}^4 + s_2^2 \theta_{01}^4)/(s_1^2+s_2^2+s_3^2))=\delta((s_1^2+s_2^2+s_3^2)/s_0)=0$
of dimension 2.

The cubic bracket is related to the quadratic one in the following way:
\begin{equation}\label{rec-3}
\{ L(u) \stackrel{\otimes}{,} L(v) \}_3 = \dfrac{1}{2} [\{ L(u) \stackrel{\otimes}{,} L(v) \}_2, L(u) \otimes \sigma_0 + \sigma_0 \otimes L(v) - s_0 \sigma_0 \otimes \sigma_0]_+,
\end{equation}
where $[A,B]_+ = AB+BA$.
Notice, that the quadratic bracket also has the similar expression:
\begin{equation}\label{rec-2}
\{ L(u) \stackrel{\otimes}{,} L(v) \}_2 = \dfrac{1}{2} [\{ L(u) \stackrel{\otimes}{,} L(v) \}_1, L(u) \otimes \sigma_0 + \sigma_0 \otimes L(v) - s_0 \sigma_0 \otimes \sigma_0]_+.
\end{equation}
It would be interesting to know the following: whether the last 2 formulas generalize to an arbitrary Lax function,
whether the whole hierarchy of symplectic structures may be obtained in this way, and what implications it has on
the integrability of Lax equations.

\section{Linear form in the rational case}

The coordinate form of Formula~(\ref{R-linear}) is
\begin{equation}\label{lin-R}
\{L_{ij}(u), L_{ls}(v)\} = \dfrac{1}{u-v} \left( (L_{lj}(u) - L_{lj}(v)) \delta_{is} + (L_{is}(v) - L_{is}(u)) \delta_{lj} \right).
\end{equation}
It is instructive to see that in the additive representation of the Lax operator
\begin{equation}\label{add-rep}
L(z) = L_0 + \sum_{i=1}^d \dfrac{a_i b_i^T}{z-z_i},
\end{equation}
Krichever-Phong's Universal Form~(\ref{KP-uform}) corresponding to $n=1$ equals
\begin{equation}\label{lin-KP2}
\omega_1 = \sum_{i=1}^d \delta a_i^T \wedge \delta b_i,
\end{equation}
and its inverse
\begin{equation}\label{lin-R2}
\{ b^l_i, a^s_j \} = \delta_{ij} \delta_{ls}, \qquad \{ a^l_i, a^s_j \} = \{ b^l_i, b^s_j \} = 0
%\{ b_i^T \stackrel{\otimes}{,} a_j \} = I \delta_{ij}, \qquad \{ a_i^T \stackrel{\otimes}{,} a_j \} = \{ b^T_i \stackrel{\otimes}{,} b_j \} = 0
\end{equation}
agrees with Brackets~(\ref{lin-R}).

Let us introduce matrices $K_0$ and $K_1$ as coefficients of the Laurent expansion of $K(z)$ at $z=\infty$
$$
K = K_0 + \dfrac{K_1}{z} + O\left( \dfrac{1}{z^2} \right).
$$
We assume that $L_0$ is in a diagonal form, i.e., $L_0=K_0$.
\begin{theorem}\label{T:lin-ratl}
Universal Form~(\ref{KP-uform}) corresponding to $n=1$ and Lax Operator~(\ref{add-rep}) is
$\omega_1 = \sum_{i=1}^d \delta a_i^T \wedge \delta b_i$.
It is a well-defined symplectic form (independent of the normalization of $\Psi$ and of gauge transformations) on the
leaves $\delta K_0 = \delta K_1 = 0$ and $\delta(b_i^T a_i)=0$.
\end{theorem}
\begin{proof}
In general, Form~(\ref{KP-uform}) depends on the normalization of $\Psi$. First, we determine conditions that make $\omega_n$
independent of the normalization, and then we compute $\omega_1$. 

A change of the normalization corresponds
to a transformation $\Psi \to \Psi V$, where $V$ is a diagonal matrix. Formula~(\ref{KP-uform}) transforms as follows:
$$
\Omega \to \Omega + 2 \text{Tr}\left( K^{1-n} \delta K \wedge \delta V V^{-1} \right).
$$
If the one-form $K^{1-n} \delta K dz$ is holomorphic on $\Gamma$, then the second term in the last formula
does not contribute to $\omega_n$. Or, equivalently,
we should restrict the one-form $k^{1-n} \delta k dz$ to some leaves where it is holomorphic.

Recall that $n=1$ under the assumptions of the theorem.
The one-form $\delta k dz$ may have poles on $\hat{\Gamma}$ above $z=\infty$ or above the points $z=z_i$ (i.e., at the poles of $k$).
Let $k_i/(z-z_i)$ be the principal part of $k(z)$ at one of its poles and
$\psi_i(z)$ be the corresponding eigenvector of $L(z)$.
Since the principal parts of both sides of the equation $L \psi = k \psi$ must be equal to each other,
we have $a_i b_i^T \psi_i(z_i) = \psi_i(z_i) k_i$. Multiplying both sides of the last equation by $b_i^T$ on the left
and dividing by $b_i^T \psi_i(z_i)$, we obtain that $k_i = b_i^T a_i$.
The one-form $\delta k dz$ is holomorphic at $z=z_i$ when its principal part vanishes, i.e., $\delta(b_i^T a_i)=0$.
Since $dz$ has a second-order pole at $z=\infty$, the one-form $\delta k dz$ is holomorphic at $\infty$ if and only if
$\delta K_0 = \delta K_1=0$.

As a by-product, it turns out that $\omega_1$ is symplectic on these leaves and does not depend on gauge transformations
$L \to g L g^{-1}$, where $g \in GL(N)$. See~\cite{IK02} for the proof when $n=1$.

Now, we evaluate~(\ref{KP-uform}).
The second term in Formula~(\ref{KP-uform}) has poles only
at $z=z_i$, $z=\infty$, branch points of $\hat{\Gamma}$, and at the poles of $\Psi(z)$.
When the eigenvectors of $L(z)$ are normalized so that the sum of their components equals one,
then $\Psi(\infty)$ is the identity matrix and $\delta \Psi(\infty) = \delta K(\infty) = 0$.
One can check that the residues of $\Omega dz$ at $z=\infty$ and at the branch points vanish.
Since the sum of all residues of a meromorphic differential must vanish, we can rewrite~(\ref{KP-uform}) as
\begin{equation}\label{lin-KP}
\omega_1 = -\dfrac{1}{2} \sum_{i=1}^d \underset{z_i}{\text{res}} \thinspace \text{Tr}\left( \Psi^{-1} \delta L \wedge \delta \Psi \right) dz.
\end{equation}
Clearly,
\begin{equation}\label{tmp1-lin}
\underset{z_i}{\text{res}} \thinspace \text{Tr}\left( \Psi^{-1} \delta L \wedge \delta \Psi \right) dz =
\text{Tr}\left( \Psi^{-1}(z_i) \delta (a_i b_i^T) \wedge \delta \Psi(z_i) \right).
\end{equation}
Only one entry of the matrix function $K(z)$ has a pole at $z=z_i$. Without loss of generality,
we assume that its principal part is $\text{diag}(\alpha_i,0,...,0)/(z-z_i)$. The identity $L \Psi = \Psi K$ implies
that
$$
a_i b_i^T \Psi(z_i) = \Psi(z_i) \text{diag}(\alpha_i,0,...,0) \text{ and }
\Psi^{-1}(z_i) a_i b_i^T = \text{diag}(\alpha_i,0,...,0) \Psi^{-1}(z_i).
$$
Consequently,
$$
b_i^T \Psi(z_i) = (\beta_i,0,...,0) \text{ and }
\Psi^{-1}(z_i) a_i = (\tilde{\beta}_i,0,...,0)^T.
$$
From the last two identities we deduce that $b_i^T a_i = \tilde{\beta}_i \beta_i$ and
$$
b_i^T \delta \Psi(z_i) \Psi^{-1}(z_i) = b_i^T \delta \ln{\beta_i} - \delta b_i^T, \qquad
\delta \Psi(z_i) \Psi^{-1}(z_i) a_i = \delta a_i - a_i \delta \ln{\tilde{\beta}_i}.
$$
Therefore, we deduce that~(\ref{tmp1-lin}) equals
$$
\underset{z_i}{\text{res}} \thinspace \text{Tr}\left( \Psi^{-1} \delta L \wedge \delta \Psi \right) dz =
2 \delta b_i^T \wedge \delta a_i,
$$
which completes the proof.

\end{proof}

We now check directly that Brackets~(\ref{lin-R2}) agree with r-matrix Poisson Brackets~(\ref{lin-R}).
Using the properties of tensor products, one can show that
\begin{equation}\label{lin-1term}
\{ a_i b_i^T \stackrel{\otimes}{,} a_i b_i^T \} = (a_i \otimes I)(I \otimes b_i^T) - (I \otimes a_i)(b_i^T \otimes I)=
(a_i b_i^T \otimes I) \mathcal{P} - (I \otimes a_i b_i^T) \mathcal{P}.
\end{equation}
Clearly, $\{ a_i b_i^T \stackrel{\otimes}{,} a_j b_j^T \} = 0$ for $i \ne j$.
Formula~(\ref{R-linear}) follows if we consider $L_0$ as a constant matrix and
use the properties of the permutation matrix $B \otimes A = \mathcal{P} (A \otimes B) \mathcal{P}$ and
$\mathcal{P}^2=I$.

A Lie algebraic interpretation of the linear brackets has been suggested in~\cite{IK02}.
Formula~(\ref{lin-KP}) may be rewritten as
$$
\omega_1 = \sum_{i=1}^d \underset{z_i}{\text{res}} \thinspace \text{Tr} \left( L \delta \Psi \Psi^{-1} \wedge \delta \Psi \Psi^{-1} \right) dz
= \sum_{m=1}^d \omega'_m.
$$
Let us define $L_m$ as
$$
L(z) = \dfrac{L_m}{z-z_m} + O(1) = \dfrac{a_m b_m^T}{z-z_m} + O(1).
$$
Then we can identify $L_m$ with a point of $\mathfrak{gl}^*(N)$
and the Lie algebra with its dual using the Killing form.
Each term $\omega'_m$ equals the Kostant-Kirillov form defined on an orbit of a co-adjoint representation
of a Lie group. As we saw before, $\delta(b_m^T a_m)=0$ on the symplectic leaves, which corresponds
to the choice of some orbit in the Lie algebra. Therefore, $\omega_1$ is the Kirillov-Kostant form on the direct product
of $d$ coadjoint orbits of $GL(N)$.

The Poisson brackets that correspond to each Kirillov-Kostant form are
$$
\{ L_m^{ij}, L_m^{ls} \} = \delta_{lj} L_m^{is} - \delta_{is} L_m^{lj},
$$
and they coincide with Formula~(\ref{lin-1term}).

\section{Quadratic form in the rational case}

As stated in the introduction, a rational matrix function $L(z)$ in general position with $d$ poles
has 2 equivalent representations:
\begin{itemize}
\item
an additive representation $L(z) = L_0 + \sum_{i=1}^d \dfrac{a_i b_i^T}{z-z_i},$ and
\item
a multiplicative representation
\[
L=
L_0
\left( I + \dfrac{p_1 q_1^T}{z-z_1} \right)
\left( I + \dfrac{p_2 q_2^T}{z-z_2} \right)..
\left( I + \dfrac{p_d q_d^T}{z-z_d} \right)=L_0 B_1 B_2 ... B_d,
\]
\end{itemize}
where $a_i$, $b_i$, $p_i$, and $q_i$ are $N$-dimensional vectors and $L_0$ is a constant matrix.

On the symplectic leaves, Formula~(\ref{KP-uform}) is invariant with respect to gauge
transformations $L \to g^{-1} L g, \Psi \to g^{-1} \Psi$,
where $g \in GL(N)$. Therefore, we may assume that the matrix $L_0$ is diagonal.

The following lemma proves the equivalency of additive and multiplicative representations for arbitrary $d$.
A similar result has been proved by Borodin in~\cite{B04}. Dzhamay~\cite{AD07} has proved the equivalence when $d=2$. 
\begin{lemma}
For any meromorphic matrix function $L(z)$ corresponding to a Zariski open subset of parameters $(z_i,a_i,b_i)$,
there exists a multiplicative representation. The converse also holds.
\end{lemma}
\begin{proof}
An additive representation follows immediately from the multiplicative one by taking the residues at the points $z_i$.

To prove the converse, we assume that we have an additive representation and construct vectors $p_i$ and $q_i$.
Let $z_1^-, z_2^-, ..., z_d^-$ be zeroes of $\text{det} \thinspace L$. Notice, for a multiplicative representation,
one has $z_i^- = z_i - p_i^T q_i$.
Let $\psi^*$ be a left eigenvector of $L$ and the corresponding eigenvalue $k$ have a pole at $z_d$.
If the principal part of $k$ is $C/(z-z_d)$, then the principal parts of both
sides of the equation $\psi^* L=k \psi^*$ are
$$
\psi^*(z_d) L_0 B_1(z_d) ... B_{d-1}(z_d) p_d \dfrac{q_d^T}{z-z_d} = \dfrac{C}{z-z_d} \psi^*(z_d).
$$
Since $\psi^*(z_d) L_0 B_1(z_d) ... B_{d-1}(z_d) p_d$ is a number, the latter equation implies that $\psi^*(z_d) \propto q_d^T$.
Likewise, if $\psi$ is the right eigenvector $L^{-1} \psi=k^{-1} \psi$, where $k$ has a zero at $z=z_d^-$, then
$\psi(z_d^-) \propto p_d$.
Since $p_d^T q_d = z_d - z_d^-$, we can recover $p_d$ and $q_d$ up to a scaling factor, which does not affect $B_d$.
We can repeat this procedure for the conjugated matrix $B_d L B_d^{-1}$ to find $B_{d-1}$.
We can find all factors $B_1, B_2, ... B_d$ in this manner.

The only thing left to prove is that $B_1 B_2 ... B_d L^{-1}$ is a constant matrix.
By construction, $L(z_d^-) p_d = 0$. By assumption, $L(z)$ is in general position, which means that
$\text{dim} \thinspace \text{ker} \thinspace L(z_d^-) = 1$ and that the residue $\text{res}_{z_d^-} L^{-1}(z) dz$ has
rank 1. Since $L L^{-1} = I$, it must be that
$$
L^{-1}(z) = \dfrac{p_d v_d^T}{z - z_d^-} + O(1)
$$
for some vector $v_d$. Since $B_d(z_d^-) p_d=0$, we conclude that the function $B_d L^{-1}$ is holomorphic at $z_d^-$.
By construction of the vector $q_d$, we have $q_d^T L^{-1}(z_d)=0$, which means that $B_d L^{-1}$ is also holomorphic at $z_d$.
Using the same arguments, we can show that
$$
B_d L^{-1}(z) = \dfrac{p_{d-1} v_{d-1}^T}{z-z_{d-1}^-} + O(1)
$$
and that $q_{d-1}^T B_d(z_{d-1}) L^{-1}(z_{d-1}) = 0$, which implies that $B_{d-1} B_d L^{-1}$ is holomorphic at $z_{d-1}^-$ and $z_{d-1}$.
By induction, we prove that $B_1 B_2 ... B_d L^{-1}$ is an entire function on the extended complex plane, hence
it has to be a constant.
\end{proof}

Now, we are in a position to prove:
\begin{theorem}\label{T:quad-rational}
Universal Form~(\ref{KP-uform}) corresponding to $n=2$ and a rational Lax matrix $L(z) = L_0 B_1 B_2 ... B_d$ in
the multiplicative representation equals
$$
\omega_2 = \sum_{i=1}^d \delta p_i^T \wedge \delta q_i.
$$
Symplectic leaves are determined by the conditions $\delta K_0 = \delta K_1 = 0$ and $\delta(q_i^T p_i)=0$,
where $K(z) = K_0 + K_1/z + O(1/z^2)$.
\end{theorem}

\begin{proof}
The proof is similar to the proof of Theorem~\ref{T:lin-ratl}. We can rewrite Formula~(\ref{KP-uform}) as
\begin{equation}
\omega_2 = -\dfrac{1}{2} \sum_{i=1}^d \underset{z_i, z_i^-}{\text{res}} \thinspace \text{Tr}\left( \Psi^{-1} L^{-1} \delta L \wedge \delta \Psi \right) dz,
\end{equation}
where $z_i^-= z_i - q_i^T p_i$ are zeroes of $\text{det} \thinspace L$.

The two-form $\omega_2$ is symplectic, independent of the normalization of $\Psi$ and of gauge transformations, provided
that the one-form $\delta \ln{K} dz$ is holomorphic on $\Gamma$ or $\delta \ln{k} dz$ is holomorphic on $\hat{\Gamma}$.
Since $dz$ has a second-order pole at $z=\infty$, we should fix $K_0$ and $K_1$, i.e., two conditions that
determine symplectic leaves for $\omega_2$ are $\delta K_0 = \delta K_1 = 0$.
Other possible singularities of $\delta \ln{k} dz$ are at the points $z_i$ and $z_i^-$.
One can check that $\delta \ln{k} dz$ is holomorphic if $\delta z_i = \delta z_i^- = 0$, which yields
the condition $\delta(q_i^T p_i)=0$.

Let us introduce matrices $T_d = L_0 B_1 B_2 ... B_d$, $T_{d-1} = B_d L_0 B_1 B_2 ... B_{d-1}$, ..., $T_1 = B_2 B_3 ... B_d L_0 B_1$.
Since $T_d \equiv L$, we have $T_d \Psi_d = \Psi_d K$ and $\Psi_d \equiv \Psi$.
Matrices $T_i$ with $i<d$ are conjugated to $T_d$, i.e., $T_i \Psi_i = \Psi_i K$, where
$\Psi_{d-1} = B_d \Psi_d$, $\Psi_{d-2} = B_{d-1} B_d \Psi_d$, ..., $\Psi_1 = B_2 B_3 ... B_d \Psi_d$.

The following transformation of $\omega_2$ is almost identical to the one used in~\cite{IK04}.
One can show that:
$$
\text{Tr}\left( \Psi_d^{-1} T_d^{-1} \delta T_d \wedge \delta \Psi_d \right) =
\sum_{k=1}^d \text{Tr}\left( \Psi_d^{-1} B_d^{-1}...B_k^{-1} \delta B_k B_{k+1}...B_d \wedge \delta \Psi_d \right)=
$$
$$
= \sum_{k=1}^d \text{Tr}\left( \Psi_k^{-1} B_k^{-1} \delta B_k \wedge \delta \Psi_k \right) -
\sum_{k=1}^{d-1} \text{Tr}\left( B_d^{-1} ... B_k^{-1} \delta B_k \wedge \delta(B_{k+1} ... B_d) \right).
$$
Notice that the last sum does not have any poles except at the points $z_i$ and $z_i^-$ and vanishes after
the summation over all residues. Therefore,
$$
\omega_2 = -\dfrac{1}{2} \sum_{i,j=1}^d \underset{z_i, z_i^-}{\text{res}} \thinspace \text{Tr}\left( \Psi_j^{-1} B_j^{-1} \delta B_j \wedge \delta \Psi_j \right) dz.
$$

The matrix $\Psi_d$ consists of normalized eigenvectors and it does not have poles at the points $z_i, z_i^-$ for any $i$
in general position.
However, matrices $\Psi_j$ may acquire poles at the points $z_i, z_i^-$ for $i > j$.
Since matrices $\Psi_j$ ($j < d$) consist of eigenvectors of $T_j$, we can normalize them: $\tilde{\Psi}_j = \Psi_j F_j$.
The matrix functions $F_j^{-1}$ are diagonal, possibly having poles at $z_i, z_i^-$ for $i > j$.
Normalized matrices $\tilde{\Psi}_j$ are holomorphic at $z_i, z_i^-$ for any $i$.

The second term on the right hand side of the identity
$$
\text{Tr}\left( \Psi_j^{-1} B_j^{-1} \delta B_j \wedge \delta \Psi_j \right) =
\text{Tr}\left( \tilde{\Psi}_j^{-1} B_j^{-1} \delta B_j \wedge \delta \tilde{\Psi}_j \right) -
\text{Tr}\left( \tilde{\Psi}_j^{-1} B_j^{-1} \delta B_j \tilde{\Psi}_j \wedge \delta \ln{F_j} \right)
$$
is holomorphic at $z_i, z_i^-$ for $i > j$, because $\delta z_i = \delta z_i^- = 0$.

Therefore, our formula for $\omega_2$ becomes:
$$
\omega_2 = -\dfrac{1}{2} \sum_{i=1}^d \underset{z_i, z_i^-}{\text{res}} \thinspace \text{Tr}\left( \Psi_i^{-1} B_i^{-1} \delta B_i \wedge \delta \Psi_i \right) dz.
$$

Plugging in the expression for $B_i$ and computing the residues, we obtain:
\begin{align}
\omega_2 = &-\dfrac{1}{2} \sum_{i=1}^d \left[ \text{Tr}\left( \Psi_i^{-1}(z_i) \left(1 - \dfrac{p_i q_i^T}{q_i^T p_i}\right)
\delta (p_i q_i^T) \wedge \delta \Psi_i(z_i) \right) + \right. \nonumber\\
&\left. +\text{Tr}\left( \Psi_i^{-1}(z_i^-) p_i q_i^T \dfrac{\delta(p_i q_i^T)}{q_i^T p_i} \wedge \delta \Psi_i(z_i^-) \right) \right]. \label{form2t}
\end{align}

Let us fix an integer $i$. Define the function $U_i$ as $T_i=U_i B_i$. The function $U_i$ is holomorphic at $z_i$, and $B_i$ has a simple
pole there. $\Psi_i$ is holomorphic at $z_i$, and $K$ is a diagonal matrix
with all but one entry holomorphic at $z_i$. Without loss of generality, assume that $K_{11}$ has
a simple pole at $z_i$. The principal part of the identity $U_i B_i \Psi_i = \Psi_i K$ implies
that $U_i(z_i) p_i q_i^T \Psi_i(z_i) \Psi_i^{-1}(z_i)$ is a diagonal matrix and
$q_i^T \Psi_i(z_i) = (\alpha_i, 0, 0, ..., 0)$, where $\alpha_i$ is some scalar function.
Taking the variation of the latter identity, we deduce:
\begin{equation}\label{eqn1}
q_i^T \delta \Psi_i(z_i) \Psi^{-1}_i(z_i) = q_i^T \delta \ln{\alpha_i} - \delta q_i^T.
\end{equation}
Similar arguments for $\Psi_i^{-1} T_i^{-1} = K^{-1} \Psi_i^{-1}$ at the point $z_i^-$ prove that
$\Psi^{-1}_i(z_i^-) p_i = (\beta_i,0,...,0)^T$ and
\begin{equation}\label{eqn2}
\delta \Psi_i(z_i^-) \Psi_i^{-1}(z_i^-) p_i = \delta p_i - p_i \delta \ln{\beta_i}.
\end{equation}
Substite~(\ref{eqn1}) and~(\ref{eqn2}) into~(\ref{form2t}) to complete the proof of the theorem.
\end{proof}
We now check that $\omega_2$ from Theorem~\ref{T:quad-rational} agrees with r-matrix Brackets~(\ref{R-quadratic}).
The Poisson brackets corresponding to $\omega_2$ are
$$
\{ p_i q_i^T \stackrel{\otimes}{,} p_i q_i^T \} = (p_i \otimes I)(I \otimes q_i^T) - (I \otimes p_i)(q_i^T \otimes I).
$$
One can check that
$$
\{ B_i(u) \stackrel{\otimes}{,} B_i(v) \} = \dfrac{(p_i \otimes I)(I \otimes q_i^T) - (I \otimes p_i)(q_i^T \otimes I)}{(u-z_i)(v-z_i)}.
$$
Consequently, we deduce:
$$
\{ B_i(u) \stackrel{\otimes}{,} B_j(v) \} = \delta_{ij} [r(u-v), B_i(u) \otimes B_j(v)].
$$
Then, if we consider $L_0$ as a constant matrix, the group property of the quadratic bracket~\cite{T87}
gives Formula~(\ref{R-quadratic}):
$$
\{ L(u) \stackrel{\otimes}{,} L(v) \} = [r(u-v), L(u) \otimes L(v)].
$$

\section{Elliptic case: general position}

Certain difficulties arise if one wants to construct a non-trivial Lax Equation~(\ref{Lax-eq}) on an elliptic curve.
In this case, the principal parts of an elliptic function $L(z)$ can't be arbitrary due to the relation
$$
\sum_i \underset{z_i}{\text{res}} \thinspace L(z) dz = 0,
$$
which is not invariant under Flow~(\ref{Lax-eq}).

Two general approaches are known to overcome this difficulty: one of them is due to Krichever and Novikov~\cite{KN80}, another one is implied
in Sklyanin's Paper~\cite{Sk82}. We consider Sklyanin's approach in Section~\ref{S:sk-approach}.
The idea of~\cite{KN80} is to introduce $N$ additional poles to the functions $L(z)$ and $P(z)$ with special dependence
on $t$, so that Equation~(\ref{Lax-eq}) is non-trivial.
Positions of the ``main'' poles and their principal parts, i.e., the set of data $(z_i, \thinspace \text{res}_{z_i} \thinspace L(z) dz)$,
determine the function $L(z)$ up to a complex scalar. In order to avoid ``pathological'' cases (e.g., when some of the poles $z_i$
coincide), we consider the values only in a Zariski open subset. We denote the divisor of the ``main'' poles by $D_+$.
In the same way as in the rational case, we assume that all poles are simple and the residues at the points $z_i$ have rank one.

The additional poles with coordinates $q_1^i, 1 \le i \le N$ play the role of so-called \emph{Tyurin parameters}
that parametrize framed stable degree $N$ holomorphic bundles. We impose the constraints:
$$
L(z) = \dfrac{\beta_i \alpha_i^T}{z-q_1^i} + L_{i1} + o(1), \qquad
\alpha_i^T \beta_i = 0, \qquad \alpha_i^T L_{i1} = \alpha_i^T \kappa_i,
$$
where the matrix $(\alpha_1, \alpha_2, ..., \alpha_N)$ is the identity matrix.
If ${\mathcal V^1}$ is a bundle that corresponds to $(q_1^i, \alpha_i)$, then $L(z)$ may be identified
with a section of $H^0(\Gamma, End({\mathcal V^1})(D_+))$, where the order of poles is bounded by the divisor $D_+$.
Lax equation~(\ref{Lax-eq}) defines an evolution of the bundle ${\mathcal V^1}$.
This construction is also applicable to Lax equations on a Riemann surface of arbitrary genus
(see~\cite{IK02} for details).

The linear symplectic form $\omega_1$ was computed in~\cite{IK02}.
In this section we compute a quadratic form for a multiplicative representation of $L(z)$.
It turns out that, in order to define a multiplicative representation, we need to introduce a sequence of vector bundles.
Suppose that we have a representation:
\begin{equation}\label{m-elliptic}
L(z) = B_d B_{d-1} ... B_1.
\end{equation}
The data $(z_i, \thinspace \text{res}_{z_i} \thinspace L(z) dz)$ depend on $2dN$ parameters. In order to completely
determine the function $L(z)$, we must specify $\sum_{i=1}^N q_1^i$. Therefore, $L(z)$ depends
on $2dN+1$ parameters. If we assume that each function $B_i$ is holomorphic at $z_j$ for $j \ne i$, then
the right hand side of~(\ref{m-elliptic}) depends on $(2N+1)d$ parameters.
However, the problem is that the Tyurin parameters for each $B_i$ are different, i.e., functions $B_i$ are endomorphisms
of different vector bundles and their composition does not make sense.

A way around this difficulty was suggested in~\cite{IK04}. The idea is to consider a sequence of
vector bundles ${\mathcal V^m}$ corresponding to the Tyurin parameters $(q_m^i, \alpha_i)$, such that ${\mathcal V}^{d+1} = {\mathcal V^1}$.
Then each $B_m \in \text{Hom}({\mathcal V^m}, {\mathcal V^{m+1}})(z_m)$ and
has $N$ poles at the points $q_m^i$, so that
\begin{equation}\label{Bm-prop1}
B_m = \dfrac{\beta_i(m) \alpha_i^T}{z-q_m^i} + O(1),
\end{equation}
where the $\beta_i(m)$ are $N$-dimensional vectors.
The inverse functions are homomorphisms of vector bundles, and
the vector bundles are in the opposite order:
$$
B_m^{-1} \in \text{Hom}({\mathcal V^{m+1}}, {\mathcal V^m})(z_m^-),
$$
where $z_m^-$ is the pole of $L^{-1}(z)$ which is distinct from the points $q_{m+1}^i$ and $D_- = \sum_m z_m^-$.
The inverse functions also have poles at the points $q_{m+1}^i$, $1 \le i \le N$, so that
\begin{equation}\label{Bm-prop2}
B_m^{-1} = \dfrac{\tilde{\beta}_i(m) \alpha_i^T}{z - q^i_{m+1}} + O(1).
\end{equation}
The function $B_m(z)$ is elliptic with a simple pole of rank one at the point $z=z_m$ and
may be written out explicitly using Weierstrass sigma functions:
$$
B_m^{ij} = f^i_m \dfrac{\sigma(z+q^i_{m+1} - q^j_m - z_m) \sigma(z-q^i_{m+1})}
{\sigma(z-z_m) \sigma(q_{m+1}^i - q_m^j) \sigma(z-q^j_m)}.
$$
Its inverse has a simple pole of rank one at the point
\begin{equation}\label{zmm}
z_m^- = z_m + \sum_{i=1}^N \left( q_m^i - q_{m+1}^i \right),
\end{equation}
where the complex numbers $q_m^i$ are subject to the periodicity conditions $q_{d+1}^i = q_1^i$.

Since the transformations $f^i_m \to \lambda_m f^i_m$ don't change $L(z)$ provided that $\prod_{m=1}^d \lambda_m=1$,
the total number of independent parameters needed to describe a chain $B_1,B_2,...,B_d$ is $2dN+1$,
which coincides with the number of parameters for the function $L(z)$.

\begin{theorem}
The quadratic symplectic form (given by~(\ref{KP-uform})) equals
\begin{equation}\label{qform-general}
\omega_2 = \sum_{m=1}^d \sum_{i=1}^N \delta \ln{\left( \dfrac{f_{m-1}^i \prod_{p \ne i} \sigma(q_m^i - q_m^p)}{\prod_{p=1}^r \sigma(q_m^i-q_{m+1}^p)} \right)} \wedge \delta  q_m^i.
\end{equation}
Symplectic leaves are determined by the conditions $\delta z_m=\delta z_m^-=0$.
Only $2d-1$ of them are independent and in general position, so the dimension of the leaves is $2d(N-1)+2$.
\end{theorem}

\begin{proof}
As opposed to the rational case, $\Omega dz$ in Formula~(\ref{KP-uform}) has poles at the points $q_1^i$.
Moreover, the residues of the two terms of $\Omega dz$ are not equal to each other.
We can rewrite~(\ref{KP-uform}) as:
\begin{equation}\label{quad-KP}
\omega_2 = -\dfrac{1}{2} \sum \text{res} \thinspace \Omega' dz
+ \dfrac{1}{2} \sum_{i=1}^N \underset{q_1^i}{\text{res}} \left( \Psi^{-1} \delta \Psi \wedge K^{-1} \delta K \right) dz,
\end{equation}
where $\Omega' = \text{Tr}\left( \Psi^{-1} L^{-1} \delta L \wedge \delta \Psi \right)$
and the first sum is taken over the points $q_1^i, z_m, z_m^-$, $1 \le i \le N$, $1 \le m \le d$.

Consider the following matrices: $T_1 \equiv L = B_d B_{d-1} ... B_1,  T_2 = B_1 B_d B_{d-1} ... B_2, ...,
T_d = B_{d-1} B_{d-2} ... B_2 B_1 B_d$. Let $\Psi_m$ be an eigen-matrix of the corresponding $T_m$, i.e.,
$T_m \Psi_m = \Psi_m K$. Matrices $\Psi_i$ with $i > 1$ are related to $\Psi_1 \equiv \Psi$ in the following way:
$\Psi_2 = B_1 \Psi_1, \Psi_3 = B_2 B_1 \Psi_1, ..., \Psi_d = B_{d-1} B_{d-2} ... B_1 \Psi_1$.

One can rewrite $\Omega'$ as:
\begin{equation}\label{quad-KP2}
\Omega' = \sum_{m=1}^d \text{Tr}\left( \Psi_m^{-1} B_m^{-1} \delta B_m \wedge \delta \Psi_m \right) -
\sum_{m=2}^d \text{Tr}\left( B_1^{-1} B_2^{-1} ... B_m^{-1} \delta B_m \wedge \delta(B_{m-1} B_{m-2} ... B_1) \right).
\end{equation}
In general position, $\Omega' dz$ does not have any residues at the points $q_m^i$ with $m > 1$. Therefore, we can safely
add them in the summation in Formula~(\ref{quad-KP}). However, the second term in~(\ref{quad-KP2}) vanishes after
this procedure. The first term in Formula~(\ref{quad-KP}) becomes:
\begin{equation}\label{quad-KP3}
-\dfrac{1}{2} \sum \text{res} \thinspace \Omega' dz = -\dfrac{1}{2} \sum_{m=1}^d \left( \text{res}_{z_m} \Omega_m dz + \text{res}_{z_m^-} \Omega_m dz +
\sum_{i=1}^N \text{res}_{q_m^i} \Omega_m dz + \sum_{i=1}^N \text{res}_{q_{m+1}^i} \Omega_m dz \right),
\end{equation}
$$
\text{ where } \Omega_m = \text{Tr}\left( \Psi_m^{-1} B_m^{-1} \delta B_m \wedge \delta \Psi_m \right).
$$
The functions $\Psi_m$ may have some poles at the points $z_i$ and $z_i^-$ with $i < m$, but
an argument identical to the one used in the proof of Theorem~\ref{T:quad-rational} shows
that they don't contribute to $\omega_2$.
Now we compute each term in Formula~(\ref{quad-KP3}):

One can check that $B_m^{-1}$ is (the computation of $\hat{L}^{-1}$ in~\cite{IK98} is almost identical to this computation):
$$
\left( B_m^{-1} \right)_{kl} = \dfrac{\sigma(z-z_m^-+q_m^k-q_{m+1}^l)\sigma(z-q_m^k)}
{f_m^l \sigma(z-z_m^-)\sigma(z-q_{m+1}^l)} \times
\dfrac{ \prod_{p=1}^N \sigma(q_{m+1}^l-q_m^p) \prod_{p \ne l} \sigma(q_m^k - q_{m+1}^p) }
{ \prod_{p \ne k} \sigma(q_m^k - q_m^p) \prod_{p \ne l} \sigma(q_{m+1}^l - q_{m+1}^p) }.
$$
Below we will use the following notation: $\text{res}_{z_m} B_m dz = P_m Q_m^T$, $\text{res}_{z_m^-} B_m^{-1} dz = \tilde{P}_m \tilde{Q}_m^T$,
where $P_m$, $Q_m$, $\tilde{P}_m$, $\tilde{Q}_m$ are the $N$-dimensional vectors:
\begin{align*}
P_m^i &= f^i_m \sigma(z_m-q^i_{m+1}),
& \tilde{P}_m^k &= \sigma(z_m^- -q_m^k) \dfrac{\prod_{p=1}^N \sigma(q_m^k - q_{m+1}^p)}{\prod_{p \ne k} \sigma(q_m^k - q_m^p)},\\
Q_m^j &= 1/\sigma(z_m-q^j_m),
& \tilde{Q}_m^l &= \dfrac{1}{f_m^l \sigma(z_m^- -q_{m+1}^l)}\dfrac{ \prod_{p=1}^N \sigma(q_{m+1}^l-q_m^p)}{\prod_{p \ne l} \sigma(q_{m+1}^l - q_{m+1}^p)}.
\end{align*}
We now compute $\text{res}_{z_m} \Omega_m dz$.
Since $T_m$ has a simple pole of rank one at $z_m$, one diagonal entry of $K$ has a simple pole at that point and
$\Psi_m$ is holomorphic at $z_m$.
The principal part of the equation $T_m \Psi_m = \Psi_m K$ implies that $Q_m^T \Psi_m(z_m) = (\varkappa_m,0,...,0)$.
Taking the variation of the last identity, we deduce that:
$$
Q_m^T \delta \Psi_m(z_m) \Psi_m^{-1}(z_m) = Q_m^T \delta \ln{\varkappa_m} - \delta Q_m^T.
$$
Using the identities $B_m^{-1}(z_m) P_m = 0$ and $Q_m^T B_m^{-1}(z_m)=0$ (which follow from $B_m B_m^{-1} = B_m^{-1} B_m = I$),
one can show that:
$$
\text{res}_{z_m} \Omega_m dz = \delta Q_m^T \wedge B_m^{-1}(z_m) \delta P_m.
$$
The identity $\Psi_m^{-1} T_m^{-1} = K^{-1} \Psi_m^{-1}$ at the point $z_m^-$ implies that
$\Psi_m^{-1}(z_m^-) \tilde{P}_m = (\kappa_m,0,...,0)^T$ and, consequently:
$$
\delta \Psi_m(z_m^-) \Psi_m^{-1}(z_m^-) \tilde{P}_m = \delta \tilde{P}_m - \tilde{P}_m \delta \ln{\kappa_m}.
$$
Using the fact that $\tilde{Q}_m^T B_m(z_m^-) = B_m(z_m^-) \tilde{P}_m = 0$, one can prove that:
$$
\text{res}_{z_m^-} \Omega_m dz = -\delta \tilde{Q}_m^T \wedge B_m(z_m^-) \delta \tilde{P}_m.
$$
Now, we compute the term $\text{res}_{q_m^i} \Omega_m dz$ in~(\ref{quad-KP3}).
We assume that the poles of $\Psi_m$ do not coincide with any of the points $q_m^i$.
Then the Taylor expansion of $\Psi_m$ is:
$$
\Psi_m = \Psi_m(q_m^i) + \Psi'_m(q_m^i) (z-q_m^i) + \dfrac{1}{2} \Psi''_m(q_m^i) (z-q_m^i)^2 + O((z-q_m^i)^3).
$$
The principal part of $T_m \Psi_m = \Psi_m K$ implies that $\alpha_i^T \Psi_m(q_m^i)=0$ since
$\Psi_m K$ is holomorphic at $q_m^i$. However, $T_m = \Psi_m K \Psi_m^{-1}$ has a pole at $q_m^i$, so
the Laurent expansion of $\Psi_m^{-1}$ is:
$$
\Psi_m^{-1} = \dfrac{\gamma_i(m) \alpha_i^T}{z - q_m^i} + \Psi_m^0 + \Psi_m^1 (z-q_m^i) + O((z-q_m^i)^2).
$$
Plugging the series of $\Psi_m$ and $\Psi_m^{-1}$ into $\Psi_m \Psi_m^{-1} = I$, we obtain:
\begin{equation}\label{rel1}
\Psi_m(q_m^i) \gamma_i(m) \alpha_i^T = 0, \quad \Psi_m(q_m^i) \Psi_m^0 + \Psi'_m(q_m^i) \gamma_i(m) \alpha_i^T = I,
\end{equation}
$$
\dfrac{1}{2} \Psi''_m(q_m^i) \gamma_i(m) \alpha_i^T + \Psi'_m(q_m^i) \Psi_m^0 + \Psi_m(q_m^i) \Psi_m^1 = 0.
$$
Formulas~(\ref{Bm-prop1}),~(\ref{Bm-prop2}) and the fact that $B_m B_m^{-1} = B_m^{-1} B_m = I$ imply that:
\begin{equation}\label{rel2}
\alpha_i^T B_m(q_{m+1}^i) = 0, \quad \alpha_i^T B_m^{-1}(q_m^i) = 0, \quad B_m^{-1}(q_m^i) \beta_i(m) = 0.
\end{equation}
Now, we show that:
\begin{equation}\label{rel3}
\alpha_i^T \left.\dfrac{\partial(B_m^{-1})}{\partial z}\right|_{q_m^i} \beta_i(m) = 1.
\end{equation}
Clearly, we have:
$$
\alpha_i^T \dfrac{\partial(B_m^{-1})}{\partial z} \beta_i(m) =
-\alpha_i^T B_m^{-1} \dfrac{\partial B_m}{\partial z} B_m^{-1} \beta_i(m).
$$
Using~(\ref{Bm-prop1}), we obtain:
$$
\alpha_i^T \left.\dfrac{\partial(B_m^{-1})}{\partial z}\right|_{q_m^i} \beta_i(m) =
\left( \alpha_i^T \left.\dfrac{\partial(B_m^{-1})}{\partial z}\right|_{q_m^i} \beta_i(m) \right)^2,
$$
so~(\ref{rel3}) follows.
Now, using Formulas~(\ref{Bm-prop1}),~(\ref{Bm-prop2}),~(\ref{rel1})-(\ref{rel3}), we compute:
\begin{equation}\label{term3}
\text{res}_{q_m^i} \Omega_m dz = -\left( \alpha_i^T \left.\dfrac{\partial(B_m^{-1})}{\partial z}\right|_{q_m^i} \delta \beta_i(m) \right) \wedge
\delta q_m^i+\alpha_i^T \Psi'_m(q_m^i) \delta \gamma_i(m) \wedge \delta q_m^i.
\end{equation}
In order to compute $\text{res}_{q_{m+1}^i} \Omega_m dz$, we need 2 additional identities that follow from $\Psi_{m+1} = B_m \Psi_m$:
\begin{equation}\label{rel4}
\Psi_m^{-1}(q_{m+1}^i) \tilde{\beta}_i(m) = \gamma_i(m+1), \quad \Psi_m(q_{m+1}^i) \gamma_i(m+1)= \tilde{\beta}_i(m).
\end{equation}
Using~(\ref{Bm-prop2}),~(\ref{rel1})-(\ref{rel2}),(\ref{rel4}), we show that:
\begin{equation}\label{term4}
\text{res}_{q_{m+1}^i} \Omega_m dz = \left( \alpha_i^T \left.\dfrac{\partial B_m}{\partial z}\right|_{q_{m+1}^i} \delta \tilde{\beta}_i(m) \right) \wedge \delta q_{m+1}^i-
\alpha_i^T \Psi'_{m+1}(q_{m+1}^i) \delta \gamma_i(m+1) \wedge \delta q_{m+1}^i.
\end{equation}
Notice that the last terms in~(\ref{term3}) and~(\ref{term4}) telescope after summing over $i$.
The only terms that we need to be careful about are those with $i=d+1$, since $\Psi_{d+1} = \Psi_1 K \ne \Psi_1$.
One can check that
$\alpha_i^T \Psi'_1(q_1^i) \delta \gamma_i(1) = \alpha_i^T \Psi'_{d+1}(q_{d+1}^i) \delta \gamma_i(d+1) +
\alpha_i^T \Psi'_1(q_1^i) \delta \ln{K(q_1^i)} \gamma_i(1)$
and
$$
\underset{q_1^i}{\text{res}} \left( \Psi_1^{-1} \delta \Psi_1 \wedge K^{-1} \delta K \right) dz =
\alpha_i^T \Psi'_1(q_1^i) \delta \ln{K(q_1^i)} \gamma_i(1) \wedge \delta q_1^i,
$$
so we finally arrive at the formula:
\begin{align}\label{qform-main}
\omega_2 = &-\dfrac{1}{2} \sum_{m=1}^d \left[ \delta Q_m^T \wedge B_m^{-1}(z_m) \delta P_m - \delta \tilde{Q}_m^T \wedge B_m(z_m^-) \delta \tilde{P}_m +
\right.\\
&\left.
+\sum_{i=1}^N \left( \alpha_i^T \left.\dfrac{\partial B_m}{\partial z}\right|_{q_{m+1}^i} \delta \tilde{\beta}_i(m) \right) \wedge \delta q_{m+1}^i
-\sum_{i=1}^N \left( \alpha_i^T \left.\dfrac{\partial(B_m^{-1})}{\partial z}\right|_{q_m^i} \delta \beta_i(m) \right) \wedge \delta q_m^i
\right].\notag
\end{align}
Straightforward, but rather lengthy computations show that:
$$
\omega_2 = \sum_{m=1}^d \sum_{i=1}^N \delta \ln{\left( \dfrac{f_{m-1}^i \prod_{p \ne i} \sigma(q_m^i - q_m^p)}{\prod_{p=1}^r \sigma(q_m^i-q_{m+1}^p)} \right)} \wedge \delta  q_m^i
$$
on the leaves, i.e., when $\delta z_m = \delta z_m^- =0$.
\end{proof}

\section{Sklyanin's case}\label{S:sk-approach}
Sklyanin~\cite{Sk82} has defined quadratic Poisson brackets on the space of Lax matrices with one pole:
\begin{equation}\label{sk-single}
L = s_0 I + \dfrac{s_1}{\pi \imath} \sigma_1 e^{\pi \imath z} \phi\left( \dfrac{\tau}{2}, z \right) +
\dfrac{s_2}{\pi \imath} \sigma_2 e^{\pi \imath z} \phi\left( \dfrac{1+\tau}{2}, z \right) +
\dfrac{s_3}{\pi \imath} \sigma_3 \phi\left( \dfrac{1}{2}, z \right),
\end{equation}
where
$$
\phi(w,z) = \dfrac{\theta_{11}(w+z | \tau) \theta'_{11}(0|\tau)}{\theta_{11}(w|\tau) \theta_{11}(z|\tau)},
$$
and $\theta_{11}(z|\tau)$ is a Jacobi theta function.
The notation for $\theta_{ij}(z)$ is the same as in Mumford~\cite{M83}. The functions $\theta_{ij}(z)$ and $\theta_{ij}(z|\tau)$ correspond to
the curve $\mathbb{C}/(\mathbb{Z}+\tau \mathbb{Z})$. The function $\theta_{ij}(z|2\tau)$ corresponds to $\mathbb{C}/(\mathbb{Z}+2\tau \mathbb{Z})$.

The function $L(z)$ has the following translational properties: $L(z+1) = \sigma_3 L(z) \sigma_3,$ and $L(z+\tau) = \sigma_1 L(z) \sigma_1,$
where $\sigma_i$ are the Pauli matrices:
$$
\sigma_1=\begin{pmatrix} 0 & 1\\1 & 0 \end{pmatrix},
\sigma_2=\begin{pmatrix} 0 & -\imath\\ \imath & 0 \end{pmatrix},
\sigma_3=\begin{pmatrix} 1 & 0\\0 & -1 \end{pmatrix}.
$$
The function $L(z)$ is elliptic on the curve $\mathbb{C}/(2\mathbb{Z}+2\tau \mathbb{Z})$, and, due to the prescribed monodromy
properties, the sum of its residues is automatically zero. This construction allows us to choose the principal part of $L(z)$
at the point $z=0$ arbitrarily provided its trace vanishes.

$L(z)$ may be identified with an endomorphism of a vector bundle on the elliptic curve $\mathbb{C}/(\mathbb{Z}+\tau \mathbb{Z})$
with a pole at $z=0$.
This bundle has degree 1 and rank 2 and is described by its section $s = (s_1(z), s_2(z))$, that transforms according to the formulas:
$$
s^T(z+1) = Q s^T(z), \text{ and } s^T(z+\tau) = \Lambda s^T(z), \text { where }
Q = \sigma_3, \text{ and }
 \Lambda = \exp{(-\pi \imath (z-u_1) - \pi \imath \tau/2)} \sigma_1.
$$
Here, we are following the notation from~\cite{Z07}. One difference with~\cite{Z07} is that we have introduced
an additional parameter $u_1$, which changes the vector bundle, but does not affect $L(z)$.
We need it to obtain proper symplectic leaves for the 2-form~(\ref{KP-uform}).

Formula~(\ref{KP-uform}) has been proved to work only in the case of degree 2 bundles,
but we can still apply it to Sklyanin's case if we consider $L(z)$ on
the curve $\mathbb{C}/(\mathbb{Z}+2 \tau \mathbb{Z})$ instead.
As a result, the degree of the corresponding vector bundle doubles.
A degree 2 bundle has 2 holomorphic sections, and one can check that $\tilde{L}(z)=g L(z) g^{-1}$ is an elliptic function
on $\mathbb{C}/(\mathbb{Z}+2 \tau \mathbb{Z})$,
where
$$
g = \begin{pmatrix} \theta_{00}(z-u_1|2 \tau) & 0\\ 0 & \theta_{10}(z-u_1|2 \tau) \end{pmatrix},
$$
and $u_1$ is an arbitrary parameter. Below we use letter $u$ instead of $u_1$ for simplicity.

The function $\tilde{L}(z)$ corresponds to a degenerate case of~(\ref{qform-general}), since its residue at $z=0$ has rank 2,
whereas all residues in the previous section had rank $1$.

\begin{theorem}
Krichever-Phong's Universal Formula~(\ref{KP-uform})
defines an hierarchy of symplectic structures on the space of matrix functions $\tilde{L}(z)$ which, in turn, depend on 5
parameters $(u,s_0,s_1,s_2,s_3)$. These structures vanish for $n > 3$, and $\omega_1,\omega_2,\omega_3$
correspond to the following Poisson brackets:
\begin{itemize}
\item A linear bracket ($n=1$):
$$
\{s_1, s_2\}_1 = -s_3, \quad \{s_1, s_3\}_1 = s_2, \quad \{s_2, s_3\}_1 = -s_1, \quad \{u, s_0\}_1 = 1,
$$
$$
\{u, s_i\} = \{s_0, s_i\}=0 \text{ for } i=1,2,3.
$$
This bracket is non-degenerate on the leaves $\delta u = \delta s_0 = \delta( s_1^2+s_2^3+s_3^2)=0$, which have dimension 2.

\item A quadratic bracket $(n=2)$:
\begin{align}\label{quad-pb}
\{s_0,s_1\}_2&=-\theta_{01}^4 s_2 s_3,
&\{s_0,s_2\}_2&=\theta_{00}^4 s_1 s_3,
&\{s_0,s_3\}_2&=-\theta_{10}^4 s_1 s_2,\notag\\
%$$
%\begin{equation}\label{quad-pb}
\{s_1, s_2\}_2 &= -s_0 s_3,
&\{s_1, s_3\}_2 &= s_0 s_2,
&\{s_2, s_3\}_2 &= -s_0 s_1,\\
%\end{equation}
%$$
\{u,s_0\}_2 &= s_0, 
&\{u,s_1\}_2 &= s_1, 
&\{u,s_2\}_2 &= s_2,\notag\\ 
&&\{u,s_3\}_2 &= s_3.&&\notag
\end{align}
The symplectic leaves for the quadratic bracket are defined by:
$$
\delta ((s_0^2 + s_1^2 \theta_{00}^4 + s_2^2 \theta_{01}^4)/(s_1^2+s_2^2+s_3^2))=0
$$
and have dimension 4.

This bracket coincides with Sklyanin Brackets~\cite{Sk82} after the symplectic reduction to submanifolds $u=0$. The submanifolds have
dimension 2 and are given by the equations
$\delta u = \delta (s_1^2+s_2^2+s_3^2) = \delta(s_0^2 + s_1^2 \theta_{00}^4 + s_2^2 \theta_{01}^4)=0.$

\item A cubic bracket $(n=3)$:
\begin{align}\label{cubic-pb}
\{s_0,s_1\}_3&=-2\theta_{01}^4 s_0 s_2 s_3, \quad
&\{s_0,s_2\}_3&=2 \theta_{00}^4 s_0 s_1 s_3,\notag\\
\{s_0,s_3\}_3&=-2\theta_{10}^4 s_0 s_1 s_2, \quad
&\{s_1, s_2\}_3 &= s_3(s_1^2 \theta_{00}^4 + s_2^2 \theta_{01}^4 - s_0^2),\\
\{s_1, s_3\}_3 &= s_2(s_0^2 - s_1^2 \theta_{10}^4 + s_3^2 \theta_{01}^4), \quad
&\{s_2, s_3\}_3 &= -s_1(s_0^2 + s_2^2 \theta_{10}^4 + s_3^2 \theta_{00}^4).\notag
\end{align}
This bracket is non-degenerate on the leaves
$$
\delta u = \delta ((s_0^2 + s_1^2 \theta_{00}^4 + s_2^2 \theta_{01}^4)/(s_1^2+s_2^2+s_3^2))=\delta((s_1^2+s_2^2+s_3^2)/s_0)=0
$$
of dimension 2.

\end{itemize}
\end{theorem}

\begin{proof}

Formula~(\ref{KP-uform}) is equivalent to:
\begin{equation}\label{simple-kp00}
\omega_n = \sum k^{1-n}(\hat{\gamma}_i) \delta k(\hat{\gamma}_i) \wedge \delta z(\hat{\gamma}_i),
\end{equation}
where $\hat{\gamma}_i$ are poles of eigenvectors $\psi$ of $\tilde{L}(z)$ on $\hat{\Gamma}$ due to the following normalization:
$\psi_1 \equiv 1$. In this case, the poles are given by the equation $\tilde{L}_{12}(z) =0$.
The symplectic leaves are determined by the condition that the one-form $k^{1-n} \delta k dz$ is holomorphic.

The proof of the theorem is a direct computation, and is similar in all three cases, i.e., when $n=1,2, \text{ or }3$.
We outline it below in the case $n=2$:

Points where the one-form $k^{-1} \delta k dz$ may fail to be holomorphic are zeroes and poles of $k$.
They correspond to zeroes and poles of $\text{det} \thinspace L(z)$.
The form will be holomorphic, if their positions are fixed, i.e., we need to impose the constraint $\delta z_0=0$, where $z_0$ is a zero
of $\text{det} \thinspace L(z)$. When $n>3$,  $\omega_n$ vanishes after we impose all necessary constraints.
The determinant of $L(z)$ equals:
$$
\text{det} \thinspace L(z) = s_0^2 + s_1^2 \theta_{00}^4 + s_2^2 \theta_{01}^4 + \dfrac{\theta_{10}^2(z)}{\theta_{11}^2(z)} \theta_{00}^2 \theta_{01}^2 (s_1^2+s_2^2+s_3^2),
$$
therefore the symplectic leaves are determined by only one condition:
\begin{equation}\label{sk-leaves}
\delta \left( \dfrac{s_0^2 + s_1^2 \theta_{00}^4 + s_2^2 \theta_{01}^4}{s_1^2+s_2^2+s_3^2} \right)=0.
\end{equation}
In the case $n=2$, Formula~(\ref{simple-kp00}) becomes:
\begin{equation}\label{simple-kp01}
\omega_2 = \sum_i \delta \ln{k(\hat{\gamma}_i)} \wedge \delta z(\hat{\gamma}_i).
\end{equation}
The spectral curve is a 2-sheeted cover of the elliptic curve $\mathbb{C}/(\mathbb{Z}+2 \tau \mathbb{Z})$
and $\psi$ has 4 poles.  Two of them are located on both sheets above the point $z=u+\tau+1/2$.
The other 2 are above the points $z=\tilde{z}$, $z=1+2\tau-\tilde{z}$ (where $L_{12}(\tilde{z})=0$)
and correspond to $k=\tilde{L}_{22}(z)=L_{22}(z)$.
Sum~(\ref{simple-kp01}) equals:
$$
\omega_2 = \delta \ln{L_{22}(\tilde{z})} \wedge \delta \tilde{z} - \delta \ln{L_{22}(1+2\tau - \tilde{z})} \wedge \delta \tilde{z} +
\delta \ln{\text{det} \thinspace L(u+\tau+1/2)} \wedge \delta u.
$$
Equation~(\ref{sk-leaves}) implies that:
$$
\delta \ln{\text{det} \thinspace L(u+\tau+1/2)} \wedge \delta u = \delta \ln{(s_1^2+s_2^2+s_3^2)} \wedge \delta u.
$$
Further computations show that:
\begin{equation}\label{KP-Sk}
\omega_2 = 2 \dfrac{s_1 \delta s_2 \wedge \delta s_3 - s_2 \delta s_1 \wedge \delta s_3 + s_3 \delta s_1 \wedge \delta s_2}{s_0 (s_1^2+s_2^2+s_3^2)} +
\delta \ln{(s_1^2+s_2^2+s_3^2)} \wedge \delta u.
\end{equation}
This form has rank $4$ and corresponds to the Poisson brackets:
$$
\{s_0,s_1\}=-\theta_{01}^4 s_2 s_3, \quad
\{s_0,s_2\}=\theta_{00}^4 s_1 s_3, \quad
\{s_0,s_3\}=-\theta_{10}^4 s_1 s_2,
$$
$$
\{s_1, s_2\} = -s_0 s_3, \quad
\{s_1, s_3\} = s_0 s_2, \quad
\{s_2, s_3\} = -s_0 s_1,
$$
$$
\{u,s_0\} = s_0, \quad
\{u,s_1\} = s_1, \quad
\{u,s_2\} = s_2, \quad
\{u,s_3\} = s_3.
$$
Notice that the direct inversion of Formula~(\ref{KP-Sk}) leads to an additional factor $1/2$ in all Poisson brackets,
e.g., the first bracket is $\{s_0,s_1\}=-\theta_{01}^4 s_2 s_3/2$. This factor appears
because we double the elliptic curve, but we omit it in all formulas for Poisson brackets.

Quadratic brackets for Sklyanin's case were also computed in~\cite{Z07}, and they coincide with those in~\cite{Sk82}.
Formula A.23 in~\cite{Z07} yields the identities:
$$
E_2\left(\dfrac{1}{2}\right) - E_2\left(\dfrac{\tau}{2}\right) = \pi^2 \theta_{00}^4, \quad
E_2\left(\dfrac{1}{2}\right) - E_2\left(\dfrac{1+\tau}{2}\right) = \pi^2 \theta_{01}^4, \quad
E_2\left(\dfrac{1+\tau}{2}\right) - E_2\left(\dfrac{\tau}{2}\right) = \pi^2 \theta_{10}^4,
$$
that allow us to simplify formulas in~\cite{Z07} to:
$$
\{s_0,s_1\}=2 \pi \theta_{01}^4 s_2 s_3, \quad
\{s_0,s_2\}=-2 \pi \theta_{00}^4 s_1 s_3, \quad
\{s_0,s_3\}=2 \pi \theta_{10}^4 s_1 s_2,
$$
$$
\{s_1, s_2\} = \dfrac{2}{\pi} s_0 s_3, \quad
\{s_1, s_3\} = -\dfrac{2}{\pi} s_0 s_2, \quad
\{s_2, s_3\} = \dfrac{2}{\pi} s_0 s_1.
$$
The latter formulas agree with ours up to a constant factor after the rescaling $s_0 \to -s_0/\pi$.

Now, we have to compare the conditions which determine the symplectic leaves in~\cite{Sk82} and in~\cite{Z07} with~(\ref{sk-leaves}).
The brackets in~\cite{Sk82} and in~\cite{Z07} have rank two, because there is no generator $u$.
The symplectic reduction of our two-form~(\ref{KP-Sk}) to the submanifolds with the constant $u$ yields
2 additional constraints (3 in total):
$$
\delta u = \delta (s_1^2+s_2^2+s_3^2) = \delta(s_0^2 + s_1^2 \theta_{00}^4 + s_2^2 \theta_{01}^4)=0.
$$
The latter formulas coincide with the conditions for the symplectic leaves in~\cite{Sk82} and in~\cite{Z07}.
\end{proof}
\emph{Remark:} The Jacobi identity for Quadratic and Cubic Brackets~(\ref{quad-pb}) and~(\ref{cubic-pb}) is equivalent
to the only relation between $\theta_{00}, \theta_{01}, \theta_{10}$, which is $\theta_{00}^4 = \theta_{01}^4+\theta_{10}^4$.
Therefore, one can get a 2-parameter family of quadratic and cubic Poisson algebras by replacing $\theta_{01}$ and $\theta_{10}$
by arbitrary complex numbers.

The proofs for Formulas~(\ref{rec-3}) and~(\ref{rec-2}) are a direct computation using Riemann's theta relations.

\section{Degree 1 bundles with an arbitrary number of poles}

Sklyanin's brackets may be generalized to the case when $L(z)$ has an arbitrary number of poles and any rank.
An explicit computation was performed in~\cite{Z07}. In this section, we introduce a multiplicative representation
for a multi-pole Lax function and show that it is natural for the quadratic brackets.
For simplicity, we consider only rank $2$ bundles.

The construction of a vector bundle is the same as in the single pole case, and
the Lax function $L(z)$ with $d$ poles has the form~\cite{Z07}:
\begin{align}\label{sk-additive}
L(z) = &\tilde{s}^0 I + \sum_{j=1}^d \left[ \tilde{s}^0_j E_1(z-z_j) I + \dfrac{\tilde{s}^1_j}{\pi \imath} \sigma_1 e^{\pi \imath (z-z_j)} \phi\left(\dfrac{\tau}{2},z-z_j\right) +\right.\\
&+\left. \dfrac{\tilde{s}^2_j}{\pi \imath} \sigma_2 e^{\pi \imath (z-z_j)} \phi\left(\dfrac{1+\tau}{2},z-z_j\right) + \dfrac{\tilde{s}^3_j}{\pi \imath} \sigma_3 \phi\left(\dfrac{1}{2},z-z_j\right) \right],\notag
\end{align}
where the parameters $\tilde{s}^0_j$ satisfy $\sum_{j=1}^d \tilde{s}^0_j = 0$, and $E_1(z) = \partial_z \ln{\theta_{11}(z|\tau)}$.

The function $L(z)$ has the following translational properties:
\begin{equation}\label{L-monod}
L(z+1) = \sigma_3 L(z) \sigma_3 \text{ and } L(z+\tau) = \sigma_1 L(z) \sigma_1.
\end{equation}
It is an elliptic function on the curve $\mathbb{C}/(2\mathbb{Z}+2\tau \mathbb{Z})$, and, due to the prescribed monodromy
properties the sum of its residues is automatically zero. This construction allows us to choose the principal parts of $L(z)$
at the points $z=z_j$ arbitrarily, provided that their traces vanish. The number of independent parameters needed to
describe each function $L(z)$ is $4d$.
In the same way as in the single pole case, the function $L(z)$ becomes elliptic
on $\Gamma=\mathbb{C}/(\mathbb{Z}+2 \tau \mathbb{Z})$ after the conjugation $\tilde{L}= g L g^{-1}$, where
$$
g(z) = \begin{pmatrix} \theta_{00}(z-u|2 \tau) & 0\\ 0 & \theta_{10}(z-u|2 \tau) \end{pmatrix}.
$$

The construction of a multiplicative representation is similar to the case of ``general position.''
We introduce a sequence of degree 1, rank 2 vector bundles $\mathcal V^m$ on $\mathbb{C}/(\mathbb{Z}+\tau \mathbb{Z})$
described by their sections $s(z)$, such that
$$
s^T(z+1) = \sigma_3 s^T(z) \text{ and } s^T(z+\tau) = \exp{(-\pi \imath (z-u_m) - \pi \imath \tau/2)} \sigma_1 s^T(z).
$$
The factors $B_i$ of a multiplicative representation are homomorphisms
$\text{Hom}({\mathcal V^i}, {\mathcal V^{i+1}})(z_i)$ with a possible pole at the point $z=z_i$.
We assume that $\mathcal V^{d+1}=\mathcal V^1$.
The following theorem relates additive and multiplicative representations:

\begin{theorem}\label{T:sk-multirep}
Function $L(z)$~(\ref{sk-additive}) has a multiplicative
representation $L(z) = B_d B_{d-1} ... B_1$ for a Zariski open subset of parameters $\tilde{s}_m^j$,
where
\begin{align*}
B_m = &s_m^0 I \phi(\Delta_m,z'_m) + \dfrac{s_m^1}{\pi \imath} \sigma_1 e^{\pi \imath z'_m} \phi\left( \dfrac{\tau}{2} + \Delta_m, z'_m \right) +\\
&+\dfrac{s_m^2}{\pi \imath} \sigma_2 e^{\pi \imath z'_m} \phi\left( \dfrac{1+\tau}{2} + \Delta_m, z'_m \right)+
\dfrac{s_m^3}{\pi \imath} \sigma_3 \phi \left( \dfrac{1}{2} + \Delta_m, z'_m \right),
\end{align*}
$\Delta_m = u_{m+1}-u_m$, $z'_m = z-z_m$, $u_1 = u_{d+1}$, and the parameter $u_1$ may be chosen arbitrarily.
\end{theorem}
\begin{proof}
The function $\text{det} \thinspace L(z)$ is elliptic on the curve $\mathbb{C}/(\mathbb{Z}+\tau \mathbb{Z})$ and in general
position it has $2d$ distinct zeroes. Let $z_1^-, z_2^-, ..., z_d^-$ be any $d$ of them and we denote the rest
with letters $\tilde{z}_1^-, \tilde{z}_2^-, ..., \tilde{z}_d^-$. Since $\text{det} \thinspace L(z)$ is an elliptic function,
it must be that
$$
2 \sum_{i=1}^d z_i = \sum_{i=1}^d (z_i^- + \tilde{z}_i^-),
$$
which makes possible to choose parameters $u_2, u_3, ..., u_d$ for any choice of $u_1$, so that $\tilde{z}_m^- = 2 z_m - 2 \Delta_m - z_m^-$ for any $m$.

Let us define the vectors $P_m, Q_m, \tilde{P}_m, \tilde{Q}_m$ as:
$$
B_m^{-1} = \dfrac{P_m Q_m^T}{z-z_m^-} + O(1), \qquad
B_m^{-1} = \dfrac{\tilde{P}_m \tilde{Q}_m^T}{z-\tilde{z}_m^-} + O(1).
$$
Then
$$
P_m Q_m^T = \left[ \left. \dfrac{d}{dz} \right|_{z_m^-} \left( det \thinspace B_m \right) \right]^{-1}
\begin{pmatrix} 1 \\ -B_m^{21}(z_m^-)/B_m^{22}(z_m^-) \end{pmatrix} \left( B_m^{22}(z_m^-) \quad -B_m^{12}(z_m^-) \right),
$$
$$
\tilde{P}_m \tilde{Q}_m^T = \left[ \left. \dfrac{d}{dz} \right|_{\tilde{z}_m^-} \left( det \thinspace B_m \right) \right]^{-1}
\begin{pmatrix} 1 \\ -B_m^{21}(\tilde{z}_m^-)/B_m^{22}(\tilde{z}_m^-) \end{pmatrix} \left( B_m^{22}(\tilde{z}_m^-) \quad -B_m^{12}(\tilde{z}_m^-) \right).
$$
Therefore, we conclude that:
$$
P_m \propto \begin{pmatrix} \pi s_m^0 \dfrac{\theta_{11}(\hat{z}_m + \Delta_m)}{\theta_{11}(\Delta_m)} + \imath s_m^3 \dfrac{\theta_{10}(\hat{z}_m + \Delta_m)}{\theta_{10}(\Delta_m)}\\
\imath s_m^1 \dfrac{\theta_{01}(\hat{z}_m + \Delta_m)}{\theta_{01}(\Delta_m)} - s_m^2 \dfrac{\theta_{00}(\hat{z}_m + \Delta_m)}{\theta_{00}(\Delta_m)} \end{pmatrix}
\text{ and }
$$
\begin{equation}\label{evectors}
\tilde{P}_m \propto \begin{pmatrix} -\pi s_m^0 \dfrac{\theta_{11}(\hat{z}_m + \Delta_m)}{\theta_{11}(\Delta_m)} + \imath s_m^3 \dfrac{\theta_{10}(\hat{z}_m + \Delta_m)}{\theta_{10}(\Delta_m)}\\
\imath s_m^1 \dfrac{\theta_{01}(\hat{z}_m + \Delta_m)}{\theta_{01}(\Delta_m)} - s_m^2 \dfrac{\theta_{00}(\hat{z}_m + \Delta_m)}{\theta_{00}(\Delta_m)} \end{pmatrix},
\end{equation}
where $\hat{z}_m=z_m^- - z_m$.

Notice that if one already has a representation $L(z) = B_d B_{d-1} ... B_1$, then
$P_1$ and $\tilde{P}_1$ span kernels of $L(z)$ at the points $z=z_1^-$ and $z=\tilde{z}_1^-$.
This is equivalent to two equations in $s_1^0,s_1^1,s_1^2,s_1^3$. Using an additional constraint
$\text{det} \thinspace B_1(z_1^-)=0$, we can recover the values $s_1^0,s_1^1,s_1^2,s_1^3$ up to a
common scalar factor for any function $L(z)$~(\ref{sk-additive}).
We apply the same procedure to the conjugated
functions $B_1 L B_1^{-1}, B_2 B_1 L B_1^{-1} B_2^{-1},...$ to construct $d$ matrix functions $B_1,B_2,...,B_d$.

Now, let us show that $L B_1^{-1} B_2^{-1} ... B_d^{-1}$ is a constant matrix proportional to $I$.
Since $L(z_1^-) P_1 = 0$, the function $L B_1^{-1}$ is holomorphic at $z=z_1^-$. Likewise, it is also holomorphic at
$z=\tilde{z}_1^-$. By construction, $B_1^{-1}(z_1)=0$, so $L B_1^{-1}$ is also holomorphic at $z_1$.
In general position, $B_1(z_2^-)$ is a non-degenerate matrix. Therefore, by the construction of $P_2$,
it must be that $L(z_2^-) B_1^{-1}(z_2^-) P_2=0$, and, consequently, $L B_1^{-1} B_2^{-1}$ is holomorphic at $z_1^-$ and $z_2^-$.
Using similar arguments and the fact that $B_m^{-1}(z_m)=0$ for all $m$, we can show that
$L B_1^{-1} B_2^{-1} ... B_d^{-1}$ is holomorphic everywhere on $\mathbb{C}/(\mathbb{Z}+\tau \mathbb{Z})$.

Each matrix function $B_m$ is elliptic on $\Gamma=\mathbb{C}/(\mathbb{Z}+2 \tau \mathbb{Z})$ after the conjugation $B_m \to g_{m+1} B_m g_m^{-1}$, where
$$
g_m(z) = \begin{pmatrix} \theta_{00}(z-u_m|2 \tau) & 0\\ 0 & \theta_{10}(z-u_m|2 \tau) \end{pmatrix}.
$$
Using this fact, one can check that the product $B_d B_{d-1} ... B_1$ satisfies~(\ref{L-monod}).
A non-degenerate holomorphic matrix function having Monodromy Properties~(\ref{L-monod}) must be proportional to $I$.
Since variables $s_m^0,s_m^1,s_m^2,s_m^3$ are defined only up to a scalar factor, we can always choose them
to make $L \equiv B_d B_{d-1} ... B_1$.
\end{proof}

The next theorem establishes a correspondence between Poisson brackets for Additive Representation~(\ref{sk-additive})
and those for a multiplicative representation. A proof using Krichever-Phong's Formula~(\ref{KP-uform})
is possible. However, we will take a different approach, since it provides a shorter proof.

\begin{theorem}
Let us define Poisson brackets on the direct product of $d$ copies of single-pole spaces
using Formulas~(\ref{quad-pb}):
\begin{align*}
\{\hat{s}_m^0,\hat{s}_m^1\}&=-\theta_{01}^4 \hat{s}_m^2 \hat{s}_m^3, 
&\{\hat{s}_m^0,\hat{s}_m^2\}&=\theta_{00}^4 \hat{s}_m^1 \hat{s}_m^3, 
&\{\hat{s}_m^0,\hat{s}_m^3\}&=-\theta_{10}^4 \hat{s}_m^1 \hat{s}_m^2,\\
\{\hat{s}_m^1, \hat{s}_m^2\} &= -\hat{s}_m^0 \hat{s}_m^3, 
&\{\hat{s}_m^1, \hat{s}_m^3\} &= \hat{s}_m^0 \hat{s}_m^2, 
&\{\hat{s}_m^2, \hat{s}_m^3\} &= -\hat{s}_m^0 \hat{s}_m^1,
\end{align*}
where we identify:
\begin{equation}\label{pb-rescale}
s_m^1 = \theta_{01}^{-1} \dfrac{\theta_{01}(\Delta_m)}{\theta_{11}(\Delta_m)} \hat{s}_m^1, \quad
s_m^2 = \theta_{00}^{-1} \dfrac{\theta_{00}(\Delta_m)}{\theta_{11}(\Delta_m)} \hat{s}_m^2, \quad
s_m^3 = \theta_{10}^{-1} \dfrac{\theta_{10}(\Delta_m)}{\theta_{11}(\Delta_m)} \hat{s}_m^3, \quad
s_m^0 = (\theta'_{11})^{-1} \hat{s}_m^0,
\end{equation}
the variables $\Delta_m = u_{m+1}-u_m$ and $u_m$ are constants from Theorem~\ref{T:sk-multirep}.
Other brackets $\{s_m^i, s_{m'}^j\}$ vanish for $m \ne m'$.

Then these brackets coincide with Quadratic Brackets~(\ref{R-quadratic}):
$$
\{ L(u) \stackrel{\otimes}{,} L(v) \} = [r(u-v), L(u) \otimes L(v)],
$$
where $r(z)$ is elliptic r-matrix~(\ref{r-elliptic}) and $L(z)=B_d B_{d-1}...B_1$.
Consequently, they also coincide with Krichever-Phong's Formula~(\ref{KP-uform}) ($n=2$) on symplectic leaves.

Symplectic leaves on the direct product are determined by $2d$ conditions
$\delta ((\hat{s}_m^0)^2 + (\hat{s}_m^1)^2 \theta_{00}^4 + (\hat{s}_m^2)^2 \theta_{01}^4)
= \delta ((\hat{s}_m^1)^2+(\hat{s}_m^2)^2+(\hat{s}_m^3)^2))=0$, and their dimension is $2d$.
\end{theorem}

\begin{proof}
Although the formula for $B_m(z)$ does not coincide with~(\ref{sk-single}) after Identification~(\ref{pb-rescale}),
one can check that
$$
\{ B_m(u) \stackrel{\otimes}{,} B_m(v) \} = [r(u-v), B_m(u) \otimes B_m(v)],
$$
where $r(z)$ is elliptic r-matrix~(\ref{r-elliptic}).
It is a direct computation using six Riemann's theta relations:
$$
\theta_{01} \theta_{01}(u') \theta_{10}(v') \theta_{10}(u'-v')=
\theta_{10} \theta_{10}(u') \theta_{01}(v') \theta_{01}(u'-v') + \theta_{00} \theta_{00}(u') \theta_{11}(v') \theta_{11}(u'-v'),
$$
$$
\theta_{00} \theta_{10}(u') \theta_{00}(v') \theta_{10}(u'-v')=
\theta_{10} \theta_{00}(u') \theta_{10}(v') \theta_{00}(u'-v') - \theta_{01} \theta_{11}(u') \theta_{01}(v') \theta_{11}(u'-v'),
$$
$$
\theta_{01} \theta_{01}(u') \theta_{00}(v') \theta_{00}(u'-v')=
\theta_{00} \theta_{00}(u') \theta_{01}(v') \theta_{01}(u'-v') + \theta_{10} \theta_{10}(u') \theta_{11}(v') \theta_{11}(u'-v'),
$$
$$
\theta_{00} \theta_{11}(u') \theta_{01}(v') \theta_{10}(u'-v')=
\theta_{10} \theta_{01}(u') \theta_{11}(v') \theta_{00}(u'-v') + \theta_{01} \theta_{10}(u') \theta_{00}(v') \theta_{11}(u'-v'),
$$
$$
\theta_{01} \theta_{11}(u') \theta_{00}(v') \theta_{10}(u'-v')=
\theta_{10} \theta_{00}(u') \theta_{11}(v') \theta_{01}(u'-v') + \theta_{00} \theta_{10}(u') \theta_{01}(v') \theta_{11}(u'-v'),
$$
$$
\theta_{01} \theta_{11}(u') \theta_{10}(v') \theta_{00}(u'-v')=
\theta_{10} \theta_{00}(u') \theta_{01}(v') \theta_{11}(u'-v') + \theta_{00} \theta_{10}(u') \theta_{11}(v') \theta_{01}(u'-v'),
$$
and six others, obtained by exchanging $u' \leftrightarrow v'$. Here $u'=u-z_m+\Delta_m$ and $v'=v-z_m+\Delta_m$.

Consequently:
$$
\{ B_i(u) \stackrel{\otimes}{,} B_j(v) \} = \delta_{ij} [r(u-v), B_i(u) \otimes B_j(v)],
$$
and the group property of the quadratic bracket~(\cite{T87}) implies:
$$
\{ L(u) \stackrel{\otimes}{,} L(v) \} = [r(u-v), L(u) \otimes L(v)].
$$
Proposition 3.3 in~\cite{HM02} and Formula~(\ref{simple-kp01}) imply that these brackets
coincide with a two-form given by Krichever-Phong's universal formula on symplectic leaves.
\end{proof}

\section{Acknowledgements.}
The author would like to thank I. Krichever, L. Takhtajan, and A. Dzhamay for interesting and helpful discussions
and suggestions.

\end{document}